# Optimal Hybrid Channel Allocation:

## Based On Machine Learning Algorithms

K Viswanadh
*Communications Research Center*
International Institute of Information technology
Gachibowli, Hyderabad-500032, India
konjeti.viswanadh@research.iiit.ac.in

Dr.G Rama Murthy
*Communications Research Center*
International Institute of Information technology
Gachibowli, Hyderabad-500032, India
rammurthy@iiit.ac.in

*Abstract*— Recent advances in cellular communication systems resulted in a huge increase in spectrum demand. To meet the requirements of the ever-growing need for spectrum, efficient utilization of the existing resources is of utmost importance. Channel Allocation, has thus become an inevitable research topic in wireless communications. In this paper, we propose an optimal channel allocation scheme, Optimal Hybrid Channel Allocation (OHCA) for an effective allocation of channels. We improvise upon the existing Fixed Channel Allocation (FCA) technique by imparting intelligence to the existing system by employing the multilayer perceptron technique.

## I. INTRODUCTION

Today's communication networks are being overwhelmed by the growth of bandwidth-intensive applications like streaming media. The industry began looking to data-optimized 4th-generation technologies, with the promise of speed improvements up to 10-fold over existing technologies. This has created a huge competition among the service-providers promising their customers offering the speed and other qualities of services. One of the most important challenges that these service providers always confront with is the, "Frequency Channel Allocation". In fact the procedures such as, frequency spectrum allocation and spectrum auction are of huge importance, when ergonomics and economic factors are taken into account.

In the age of "spectrum drought", spectrum is regarded as one of the most important natural resources. With the number of users using smart devices exploding alongside with the emergence of concepts like "Internet of Things", the spectrum is getting heavily crowded and hence the need for utilizing the spectrum resources is of utmost importance. Frequency reuse is the basic strategy that has been employed over years to conserve frequency [1].

## II. PREVIOUS WORK

A given radio spectrum (or bandwidth) can be divided into a set of disjoint or non-interfering radio channels by exploiting the physical characteristics of a radio wave. In cellular mobile communication systems design, the decision on channel assignment schemes is crucial [1]. A channel assignment scheme is chosen based on the factors like received signal strength, carrier-to-interference ratio, co-channel reuse distance. The objective of any channel assignment scheme would be to enhance the received signal strength, minimize the carrier-to-interference ratio and decrease the co-channel reuse distance.

There are many ways in which the channel allocation schemes can be classified [2]. Based on the relation between the number of cells in a cluster and the number of allocated channels, channel allocation schemes are classified as follows:

- Fixed Channel Allocation
- Dynamic Channel Allocation
- Hybrid Channel Allocation

**Fixed Channel Allocation (FCA):**
In this scheme, the number of channels that are allocated to a cell is fixed and is determined on the basis of co-channel reuse factors. Further we can classify FCA into uniform channel allocation and non-uniform channel allocation. As the name suggests, the number of channels assigned to each cell is uniform in channel allocation scheme [3]. But the temporal and spatial variations in traffic across the cell make the uniform channel scheme a ineffective one. So generally a non-uniform channel allocation scheme with different borrowing schemes would be employed [4].

**Dynamic Channel Allocation (DCA):**
In this scheme, unlike the previous scheme, all the channels are placed in a central pool rather than hard-assigning of channels to each cell. The channels placed in the central pool are assigned dynamically to radio cells in accordance with the new calls that arrive to the system. After a call is completed, its channel is returned to the central pool [5]. Again based on the practical feasibilities and requirements, DCA can be further implemented in two ways: Centralized DCA and Distributed DCA. In centralized DCA, there would be a centralized controller that manages all the channel assignment operations. But this would lead to huge centralized overhead because of huge number of hand-shaking protocols that have to be followed before a link gets established. To avoid the centralized overhead, one can use the distributed DCA.

**Hybrid Channel Allocation (HCA):**
As the name suggests, this scheme employs the combination of both, fixed and dynamic channel allocation techniques. Out of the total number of available channels, some number of channels would be assigned to each cell in a manner similar to that of FCA, but the remaining number of channels would be placed in a channel pool and would be assigned to the cell based on the traffic, bandwidth and other factors [6].

### III. NOVEL APPROACHES TOWARDS OPTIMAL HCA

In this section, a novel concept in FCA, as "**Intelligent FCA**" is introduced and discussed in detail. Later in this research paper, a system employing the combination of Intelligent FCA and DCA schemes is reasoned to be **optimal HCA.**

#### A. *Intelligent FCA: Concept*

In the traditional HCA, the total number of channels will be divided into 3:1 ratio for fixed and dynamic channels respectively. One can infer that, the percentage of fixed channels clearly dominates the number of dynamic channels. Therefore it is important to focus on the technique which allocates the fixed channels such that the channels can be used more effectively.

- In this paper, we propose a new technique, "**intelligent FCA**", employing which a sense of cognition is applied in the allocation of the fixed channels.

In **Intelligent Fixed Channel Allocation** scheme, unlike the traditional approach, machine learning techniques are employed to study the traffic patterns of each cell. Based on the obtained traffic patterns, the channel allocation is made. This is different from the conventional FCA in a way that the number of channels in the conventional FCA is fixed and equal, while in intelligent FCA, the number of channels is fixed but not equal. This section describes two different approaches to implement intelligent FCA.

Before we discuss the schemes in detail, firstly we define two crucial parameters, "Idle Time" and "Packet Count" that will affect the number of channels that are to be allocated.

Throughout the paper, we consider 'M' base stations namely $B_1, B_2, ...., B_M$, the **Idle Time** and **Packet Count** are stated as $t_1, t_2, ...., t_M$ and $D_1, D_2, ...., D_M$ respectively and the corresponding probabilities be $P(t_i)$ and $P(D_i)$ respectively.

**Idle Time**: Idle Time is defined as the total amount of time, for which a base station remains idle in a given time interval.

**Probability of Idle Time:** Probability of **idle time** of $i^{th}$ base station is the ratio of idle time of that $i^{th}$ base station to the sum of idle times of all the base stations and is given by:

$$p(t_i) = \frac{t_i}{\sum_{i=1}^{M} t_i} \quad (1)$$

**Packet Count:** Packet Count is the total number of packets transferred by a base station in a given time interval.

**Probability of inverse Packet Count:** Probability of inverse **packet count** of $i^{th}$ base station is the ratio of the inverse of Packet Count that $i^{th}$ base station to the sum of Packet Count of all the base stations and is given by:

$$p(D_i) = \frac{\frac{1}{D_i}}{\sum_{i=1}^{M} \frac{1}{D_i}} \quad (2)$$

Here the values of **Idle Time** and **Packet Count** are evaluated from the machine learning analysis. After the evaluation, the probabilities of **Idle Time** and **Packet Count** can be calculated. These probabilities have a unique relation with number of channels allotted ($n_i$) to the cell (or Base Station). The relation is as follows,

[1]. As the probability of **Idle Time** for a Base Station increases, the required number of channels for that Base Station decreases.

$$n_i \; \alpha \; \frac{1}{P(t_i)}$$

[2]. As the probability of inverse of **Packet Count** for a Base Station increases, the required number of channels for that Base Station decreases.

$$n_i \; \alpha \; \frac{1}{P(D_i)}$$

We can infer that the number of channels required for a Base Station is inversely proportional to the probability of either **Idle Time** or **Packet Count**.

**Note:** Throughout this paper, the term "probability of base station" refers to any one of the above probabilities.

The following equation represents the average number of channels allocated to the each cell (or Base Station),

$$\sum_{i=1}^{M} n_i p_i \quad (3)$$

Subject to the constraint, $\sum_{i=1}^{M} n_i = L$.

For the effective utilization of resources (channels), the above value should be as minimal as possible. So our objective is to calculate the number of channels ($n_i$) such that the value depicted in equation (3) is as minimal as possible.

To compute the number of channels ($n_i$) that reduce the average, we propose the following methods:
1) "**Integer Linear Programming**" based technique and
2) "**Source Coding**" based technique

*1) "Integer Linear Programming" based technique:*

- In the integer linear programming, the probabilities of each cell are to be sorted in descending order. Then the assigned number of channels of each cell has to be in ascending order to minimize the equation (3). As the increase of number of channels depends on the manner in which probabilities decrease.

If we consider the number of channels to be in Arithmetic Progression, (satisfying the above condition), we consider the following three cases:

**Case 1**: In this case, we will assign an arithmetic progression numbers from 1 to M to the base stations in the descending order of their probabilities (simplest way with less complexity). Then the actual number of channels allocated to a base station is the constant 'c' times to the 1 to M. The scaling factor is given by,

$$c = \frac{L}{\frac{M^2 + M}{2}} \quad (4)$$

From the equation (4) the assigned channels to the cells becomes c, 2c, 3c, 4c,….., Mc.

**Case 2**: In this case, we assign the AP series, "a, a+d, a+2d, ….., a+(M-1)d" number of channels to each cell, where a and d are the required variables, that are to be computed. The upper bound ($L_{max}$) and lower bound ($L_{min}$) on the allocable number of channels to any base station, can be determined from the Machine Learning Techniques.

When the difference between $L_{max}$ and $L_{min}$ is greater than the total number of cells (i.e. M), there would be a need for optimizing the allocation scheme since the distribution of number of channels to each cell would be non-uniform. This is not the case, when the difference is minimal as the channel distribution would be far more uniform than the previous case.

From the above statements, we can make a valid assumption that
$$a = L_{min}$$
Then the integer variable d would be

$$d = \frac{L_{max} - L_{min}}{M}.$$

(Note that the right hand side is a non-integer value, would be rounded to the previous integer value since d has to be an integer because in this case it represents number of channels allocated to a base station).

**Case 3**: In this case we employ **Linear-Diophantine equation based technique [7]** to get the parameters for channel assignment. Here also we consider an AP series "a, a+d, a+2d, ….., a+(M-1)d" as number of channels to each cell. Now the following procedure is employed to obtain 'a' and 'd'.
We know that, total number of channels of each cell adds upto the available number of channels.
$$a + (a+d) + (a+2*d) + ...... + (a+(M-1)*d) = L$$

LHS is in A.P. therefore,
$$M*a + (\frac{M(M-1)}{2})*d = L \quad (5)$$

The equation (5) is in the form of Linear-Diophantine equation $ex + fy = g$ where e=M and f=$\frac{M(M-1)}{2}$ and g=L.

So, in accordance with Linear-Diophantine equation, for equation (5) to have a solution (probably infinitely many solutions), L should be an integer multiple of GCD of constants in equation (5) (i.e. GCD of M and $\frac{M(M-1)}{2}$ which is 'M' for 'odd' value of M. For 'even' value of M, we can find GCD using Euclidean algorithm [8]).

$$\therefore L = kM \quad (6)$$

Here '$k$' is an integer value. So L will be chosen in such a way that the equation (6) holds true.
Now, from the solution of Linear-Diophantine equation the variables 'a' and 'd' assumes infinitely many solutions. But out of them only finitely many set of solutions are of interest to us. To find the unique set of values for 'a' and 'd', we consider all the solutions of 'a' and then compare with $L_{min}$, then assign the value to 'a'(i.e. from the set of solutions), which is nearer to $L_{min}$.

**Case 4:** In this case we assume Geometrical Progression (GP) series as the number of channels required by each cell. The GP series is $a, ar, ar^2, \ldots, ar^{M-1}$.
This case is only considered when the following expression satisfies,

$$L_{max} - L_{min} > L_{min}(r^{M-1} - 1) \quad (7)$$

The value of 'r' in equation (7) should be '2' or more. Now we assign a= $L_{min}$ and 'r' should be maximum integer value which satisfies equation (7).

*2) "Source Coding" based technique:*

In this technique, the traditional "source coding" approach is applied to the channel allocation schemes, i.e. minimize $\sum_{i=1}^{M} n_i p_i$ subject to Kraft's inequality $\sum_{i=1}^{M} D^{-n_i} \leq 1$. The number of allocable channels is determined from ratio of probabilities of each cell.

Huffman coding is one of the prominent techniques employed in source coding. In Huffman coding, the obtained code lengths are Kraft numbers, i.e. they satisfy Kraft's inequality [9]. So here we employ the concept of Huffman Coding in the channel allocation schemes. Firstly the probabilities of all the base stations are made known to every other base station. Then the process of Huffman Coding takes place at each base station. Then each base station sends its code word (which indicates number of channels assigned to each cell) to the central controller which handles the channel assignment schemes. So here, we can say that these number of channels allocated to the each cell would be a Kraft Number.

If we assert the number of channels in a base station to be a Kraft Number, then the average number of channels in a $i^{th}$ base station is given by,

$$n_i = \log_2 P_i$$

As the number of channels allocated to a base station be $n_1, n_2, n_3, \ldots, n_M$, then the ratios of the number of channels required for each base station is evaluated as follows,

$$\begin{aligned} n_1 : n_2 : \ldots : n_M = \\ (-\log_2 P_1) : (-\log_2 P_2) : \ldots : (-\log_2 P_M) \end{aligned} \quad (8)$$

But $n_1 + n_2 + n_3 + \ldots + n_M = L \quad (9)$

From equations (8) and (9), we can compute the values of $n_1, n_2, n_3, \ldots, n_M$.

Thus, the concept of Optimal HCA is proposed by allocating channels to the each cell from the total number of available channels in the Fixed Channels division. (Determined from any of the above algorithms) If the channel requirement in a particular cell is more than the predicted (i.e. more than the allocated channels) then channels from the Dynamic pool will allocated to that cell.

Along the lines of above discussion, it is possible to consider average number of packets received as the criteria for optimization. With such criteria if the normalized probability in a base station is high the number of channels allocated should be high. Thus we are interested in maximization problem (formulated as an integer linear programming) and similar results as above can apply.

### IV. MACHINE LEARNING ALGORITHMS

In this paper, the purpose of employing the machine learning is to compute the crucial parameters such as $(t_i, D_i, L_{max}, L_{min})$ which are used in section-III. Techniques such as Multilayer Perceptron (MLP) can be employed to calculate the required parameters. Multilayer Perceptron is a feed forward artificial neural network model that maps sets of input data onto a set of appropriate outputs. MLP is proved to be effective among other conventional data traffic prediction techniques such as ARIMA, fractional ARIMA [10].

To compute the parameters we feed the following information to the MLP algorithm.
1. Base Station Number (BSN)
   This is in between 1 to M. as mentioned above.
2. Day (D)
   This indicates whether it is working day or weekend.
3. Slot in a day (SD)
   This indicates at what period of time in a day we are observing the parameters.

The following figures depicts the MLP algorithm,

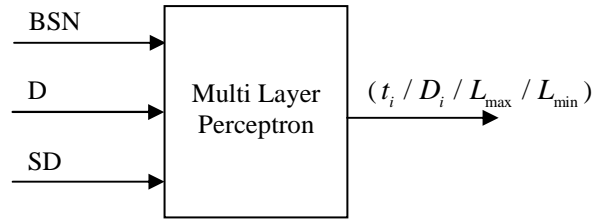

Figure-1: Representing MLP

### V. CONCLUSION

The problem of efficient channel allocation or spectrum usage can be achieved by efficient channel allocation schemes rather than opting for the age-old techniques. This can be achieved by studying the pattern of traffic at each base station and by imparting a sense of cognition to channel assignment controller.

ACKNOWLEDGEMENT

The authors would like to thank Andrew Odlyzko of University of Minnesota for discussion related to section III.